"Probing the energy conversion pathways between light, carriers and lattice in real time with attosecond core-level spectroscopy"

T.P.H. Sidiropoulos[1,†], N. Di Palo[1,†], D.E. Rivas[1,2,†], S. Severino[1], M. Reduzzi[1], B. Nandy[1], B. Bauerhenne[3], S. Krylow[3], T. Vasileiadis[4], T. Danz[5], P. Elliott[6,7], S. Sharma[6], K. Dewhurst[7], C. Ropers[5], Y. Joly[8], K. M. E. Garcia[3], M. Wolf[4], R. Ernstorfer[4], J. Biegert[1,4,9,*]

[1]ICFO - Institut de Ciencies Fotoniques, The Barcelona Institute of Science and Technology, 08860 Castelldefels (Barcelona), Spain
[2]European XFEL GmbH, Holzkoppel 4, 22869 Schenefeld, Germany
[3]Theoretische Physik, FB-10, Universität Kassel, 34132 Kassel, Germany
[4]Fritz Haber Institute of the Max Planck Society, Berlin, Germany
[5]4th Physical Institute - Solids and Nanostructures, University of Göttingen, Germany
[6]Max-Born-Institut für Nichtlineare Optik und Kurzzeitspektroskopie, 12489 Berlin, Germany
[7]Max-Planck-Institut für Mikrostrukturphysik, Weinberg 2, 06120 Halle, Germany
[8]Université Grenoble Alpes, CNRS, Grenoble INP, Institut Néel, 38000 Grenoble, France
[9]ICREA - Institució Catalana de Recerca i Estudis Avançats, Barcelona, Spain
*Correspondence to: jens.biegert@icfo.eu
†These authors contributed equally

Detection of the energy conversion pathways, between photons, charge carriers, and the lattice is of fundamental importance to understand fundamental physics and to advance materials and devices. Yet, such insight remains incomplete due to experimental challenges in disentangling the various signatures on overlapping time scales. Here, we show that attosecond core-level X-ray spectroscopy can identify these interactions with attosecond precision and across a picosecond range. We demonstrate this methodology on graphite since its investigation is complicated by a variety of mechanisms occurring across a wide range of temporal scales. Our methodology reveals, through the simultaneous real-time detection of electrons and holes, the different dephasing mechanisms for each carrier type dependent on excitation with few-cycle-duration light fields. These results demonstrate the general ability of our methodology to detect and distinguish the various dynamic contributions to the flow of energy inside materials on their native time scales.

Light-matter interaction, i.e. the interaction between light, carriers, and lattice[1], is fundamental to understand nature and to advance modern society. The detection is still a challenge since the various interactions occur on time scales which can be ultrafast, i.e., they occur within attoseconds, and range into picoseconds; the interaction between electrons, holes and lattice is not rigorously separable by different temporal scales, and the interaction is often correlated due to the underlying many-body quantum physics; see Fig. 1a. Without discrimination of the various mechanisms, however, it is difficult to understand when and why an excitation manifests adversely, for instance, as limited transport of charge carriers in metals, or as a reduced exciton lifetime in organic solids. Conversely, it is important to understand why an excitation causes mutual attraction of Fermion pairs[2], Bosonic qubits [3], or how Floquet states can potentially stabilize high-temperature superconductivity [4] and induce topological quantum phases [5]. To address such canonical problems, methods are sought which provide a holistic view over how a light field creates a non-equilibrium state, how the state evolves into the multi-body state of excited electrons and holes, whether exciton pairs are stable, and which excitation of phonons leads to the relaxation of carriers and the lattice. Amongst prominent methods to address this challenge, photoemission spectroscopies provide a detailed momentum-resolved insight into the electronic structure, but the probing depth in a solid is very limited and information about the lattice [6] is inferred indirectly, or with the aid of theoretical modeling [7]. In contrast, lattice dynamics are readily investigated with inelastic scattering methods such as Raman [8,9], neutron, electron or X-ray scattering[10] and electron energy loss spectroscopy [11], but those methods only yield indirect information on the electronic structure. Evidently, each of these powerful methods has its own merits. But, to unravel the various, often intricated, contributions [12,13] of carriers and phonons in solids [14,15], it is challenging, and many times even impossible, to combine the information from different methods to gain an understanding of the physics. It is thus highly desirable to extract carrier and lattice information from a single measurement [16,17].

Here, we show that core-level X-ray absorption near edge structure (XANES) spectroscopy with attosecond soft X-ray (SXR) pulses [18,19] meets the challenge in a single experimental method. The XANES method is based on the absorption of an X-ray photon whenever a dipole-allowed transition of a core-level electron is possible, either to a bound or to a continuum state. Core-level XANES is element selective and orbital specific, and thus it is a sensitive probe of electronic structure. It is important to realize that unambiguous mapping of the XANES absorbance to the material's density of states (DOS) is strictly ensured only with *K-shell* XANES since the 1s core transition does not suffer from final state multiplet effects[20]. This is distinct from measurements with XUV (attosecond) pulses, which access transitions from higher-lying core or inner-valence states, and, whose interpretation depends on the exact case and may require advanced theory. Figure 1b illustrates this point by showing that the absorbance from K-shell XANES unambiguously matches with the DOS calculated with density functional theory. Our conceptual approach is thus to time-resolve the change in electronic structure of a material with attosecond K-shell XANES. To realize such measurement, it is vital to combine high-energy resolution with coverage of an extensive energy range of tens of eV to disentangle the various contributions to electronic structure change. Figure 1a shows this concept and indicates how the simultaneous measurement of energetically distinct signatures in the time-resolved XANES spectrum allows distinguishing electron, hole and lattice dynamics. For instance, the attosecond time resolution allows detecting the buildup of coherences (polarization) even during the excitation light field's oscillation at THz to PHz oscillation frequency. The spectral range within a few eV of the Fermi level records changes in electronic occupation. At the same time, alteration of the electronic structure (also) far from the Fermi level records nuclear motion.

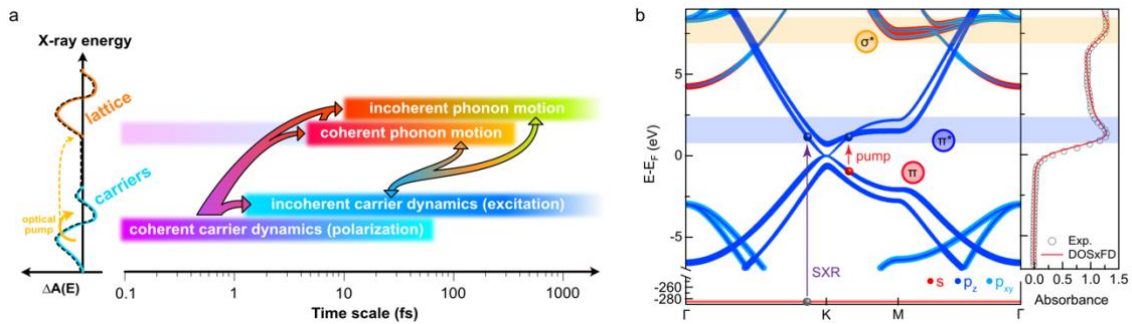

**Fig. 1 Carrier and phonon dynamics in graphite.** (a) The left panel indicates two areas which are separated in x-ray photon energy, where changes in the differential absorption occur due to carrier and lattice motion. The right panel indicates the vast time-overlapping physical effects and their interplay. (b) State-resolved band structure calculated in the Local Density Approximation (LDA) with an ultra-soft Vanderbilt potential and Perdew-Burke-Ernzerhof (PBE) exchange correlation with Quantum Espresso[21]. The blue and orange shaded areas indicate $\pi^*$ and $\sigma^*$ bands originating from $p_z$ and $sp_2$-hybridized states. The arrows show electronic transitions through SXR (violet) and near-infrared (NIR, red) pulses. The measured XANES spectrum (black circles), under 40° incidence angle, is shown on the right. Electronic transitions from the carbon 1s state (bound by 284.2 eV) to unoccupied states in both the $\pi^*$ and the $\sigma^*$ bands (1.5 and 7 eV above the Fermi level) dominate the spectrum. The XANES spectrum is in excellent agreement

with a numerical model (red) based on the density of states (DOS) and a Fermi-Dirac distribution at room temperature (see SI for further details).

We apply our methodology to graphite since the material is extensively studied and serves as a challenging benchmark due to the overlapping time scales of ultrafast carrier-carrier (c-c) and carrier-phonon (c-ph) scattering; i.e., on time scales of 5 to 10 fs [22,23] (for c-c) and up to 100 fs (for c-ph) [24–28]. Despite the plethora of existing investigations on graphite, the exact mechanism behind electron-phonon coupling (EPC), and the involvement of the various phonon modes together with their time evolution is unclear. The role of the two prominent (strongly-coupled) phonons, the $A_1'$ and $E_{2g}$ phonons, their real-time evolution, and relative contribution to EPC has never been measured. Identification of the loss channel and the nature of de-excitation is especially important when considering that EPC is the microscopic origin for 1/f noise[29] in electronic circuits, and the main reason for decoherence in quantum computing[30,31]. Similarly, the scattering mechanisms that photo-excited carriers undergo in bulk graphite are still under debate[22,32] despite the prominence of graphite in devices for energy transfer and storage.

To address these open points for graphite, we start our investigation with the (un-pumped) XANES of 95-nm free-standing graphite with a 165-as FWHM SXR probe pulse whose spectrum ranges from 250 to 500 eV [33–35]. This probe accesses the K-edge core-to-valence transition from the carbon 1s state (bound by 284.2 eV) and the broad spectral coverage ensures probing around, and far above, the Fermi level at once. Figure 1b shows the calculated bandstructure together with its orbital character. We orient the material's basal plane at 40° with respect to the linearly polarized SXR pulse to probe a wide range of orbital character[34]. This allows to interrogate the $p_z$ orbitals, normal to the Basal plane, which constitute the ($\pi$) valence band (VB) and ($\pi^*$) conduction band (CB). In contrast, sp2-hybridized orbitals ($\sigma^*$) lie inside the basal plane and are thus very sensitive to in-plane phonon motion. Figure 1b (right) shows the excellent agreement between the calculated density of states (DOS) multiplied by the Fermi Dirac (FD) function at room temperature with the measured static absorbance. The $\pi^*$ and $\sigma^*$ antibonding states, 1.5 and 7 eV above the Fermi energy are readily identified with the $\overline{M}$- and $\overline{\Gamma}$-points in the Brillouin zone. A Fermi-Dirac fit to the measured absorbance yields that the material is n-doped with a chemical potential of $200\pm45$ meV.

To study the dynamic flow of energy inside graphite, we photo-excited the material with an ultrashort pump pulse and varied its fluence and photon energy for a range of measurements. We used ultrashort $11 \pm 1$ fs pump pulses at a photon energy of 0.7 eV, and $15 \pm 1$ fs at 1.6 eV to induce a $\pi - \pi^*$ transition at the vicinity of the $\overline{K}$-point (see Fig. 1b). We varied the pump fluence between $2.8 \pm 0.2$ mJ/cm$^2$ and $81 \pm 5$ mJ/cm$^2$ to investigate different regimes of carrier dynamics; see the SI for the complete set of data. Figure 2a shows the differential absorption ΔA(*E*), which is the difference between the measured absorption of the pumped and of the un-pumped material, for the highest fluence of 81 mJ/cm$^2$ and with 0.7 eV photon energy.

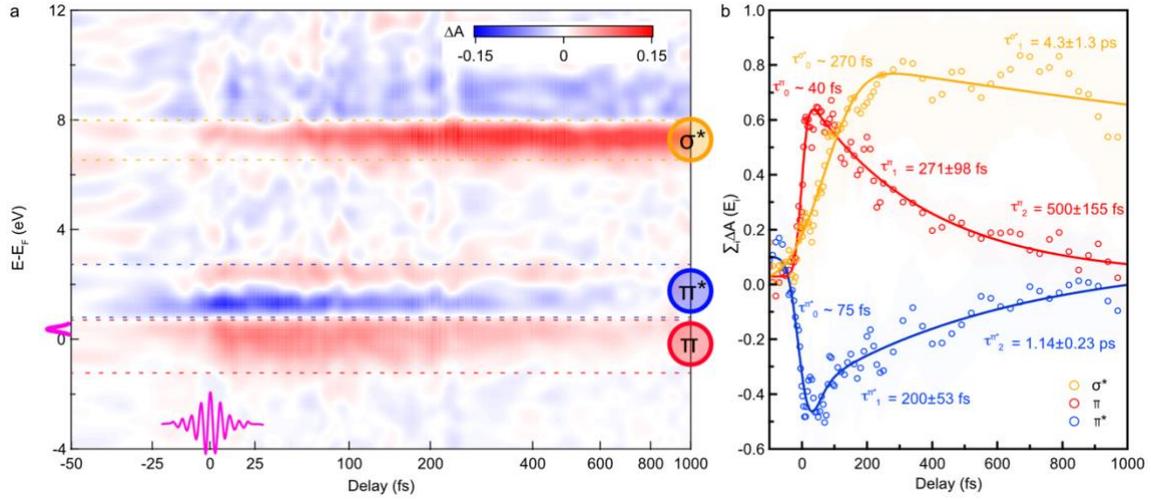

**Fig. 2 Time-dependent XANES measurement.** (a) Shown is the differential absorption ΔA(*E*) between the XANES spectrum with and without pump pulse. The observed features are identified as arising from electrons at the bottom of the CB ($\pi^*$), holes at the top of the valence band ($\pi$), and predominately from optical phonons ($\sigma^*$); note that the material is n-doped to 200 meV. The pump pulse bandwidth and duration are indicated in pink. (b) Sum of the differential absorption $\sum_i \Delta A(E_i)$ over the respective energy range for $\pi$ (red), $\pi^*$ (blue), and $\sigma^*$ (yellow). The curves are fitted with a double exponential convolved with a Gaussian.

The pump pulse pulse and its spectrum are indicated in pink in Figure 2a. Immediately apparent in Fig. 2a are changes of up to 15% in differential absorption, both positive (red) and negative (blue). Figure 1b allows to identify these features as $\pi$ bonding state (~-0.5 eV), and as $\pi^*$ (~ 2 eV) and $\sigma^*$ (~7.5 eV) antibonding states.

To remove pump-induced broadening and shifting contributions to the line shape, we analyze the absorption spectra at different temporal delays, analog to the procedure detailed in Ref. [36]; see the SI for details. This procedure reveals that the signal between -1.5 eV to 3 eV originates indeed from occupied (electron) $\pi^*$-character CB states and from unoccupied (hole) $\pi$-character VB states. The absorption change between -1.5 eV to 3 eV is energetically much broader than the bandwidth of the pump pulse, but this is explained by the high carrier concentration of $5\times10^{22}$ cm$^{-3}$ at the high pump fluence shown in Fig. 2a. We further find that the signal between 6 and 10 eV originates from a shift and broadening of the $\sigma^*$ state at ~7.5 eV. Having determined the position of the respective states, we fit a Gaussian function with bi-exponential decay along their energetic positions. Figure 2b shows the result which reveals that the $\pi^*$ (electron) signal rises more slowly ($\tau_r^e = 75 \pm 5$ fs) than the $\pi$ (hole) signal ($\tau_r^h = 40 \pm 5$ fs). Conversely, electrons relax faster ($\tau_1^e = 200 \pm 53$ fs) than holes ($\tau_1^h = 271 \pm 98$ fs). Interestingly, all carrier time scales are much longer than the C1s core-hole lifetime in graphite (1.6 fs [37]), which attributes the measured behavior to Debye screening and Landau damping of charge carriers [38]. In contrast, the $\sigma^*$ signal exhibits a markedly slower dynamics with a rise time of $\tau_0^{\sigma^*} = 270 \pm 10$ fs and decay time of $\tau_1^{\sigma^*} = 4.3 \pm 1.3$ ps. This is in accord with expected time evolution of phonon motion.

We now detail the early time dynamics for excitation at a much lower fluence of 2.8 mJ/cm$^2$ for a photon energy of 1.6 eV and contrast the findings against a similar fluence of 3.2 mJ/cm$^2$ but with photon energy of 0.7 eV; see SI for XANES data. Figure 3a and b show results of the attosecond-resolved measurement with a pump-probe delay step size of 0.6 fs. Immediately apparent in Fig. 3a and b is the buildup of coherent charge oscillations, i.e., polarization of the material. These oscillations occur at occupied states below (orange) and unoccupied states above (blue) the Fermi level. A Fourier analysis, shown in the SI, reveals that the oscillations occur predominantly at the pump carrier frequency; carrier periods are indicated by the vertical dashed lines in Fig. 3a and b. We attribute the observed charge oscillations at the fundamental rather than second harmonic of the optical pump frequency (as observed in attosecond-resolved experiments in non-Dirac materials[39–42]) to the prevalent excursion of carriers in the near-linear potential in close proximity of the K-point[43], thus leading to a linear response at the fundamental driving frequency. A low-pass filtered fit through the data (blue and orange solid lines) reveals the incoherent background, which is due to the dephasing of coherent charge oscillation. We observe the concomittent incoherent background raising within a few oscillations of the light field, signifying the ultrafast transfer of energy from the light field into the electron and hole excitation of the material.

An important question that we addressed next is the nature of the observed ultrafast carrier-carrier interaction, during excitation and its dephasing. Our aim is to distinguish between the pertinent mechanisms of Impact Excitation (IE) and Auger Heating (AH)[44] and to determine whether both carrier types, electrons and holes, are exposed to identical or different scattering mechanisms. Such an investigation is especially interesting when considering that IE leads to a multiplication of carriers, while AH results in a decrease in the number of carriers. A dynamic imbalance between electrons and holes will leads to a changing chemical potential and varying probability for carrier recombination. Figure 3e depicts how a simple comparison of trends for the two quantities IE and AH [45,46] allows distinguishing between the mechanisms [44] directly from the measurement. To capture the excitation and the decay of the incoherent background, we conducted additional measurements over a longer time range, up to 90 fs with time steps of 3 fs. Figures 3c and d show the result for which we plot the number of carriers $N_c = \sum_i |\Delta A(E_i)|$ together with their occupation-normalized mean (kinetic) energy $\langle E_c \rangle = \sum_i E_i \, \Delta A(E_i) / N_c$. Examination of the low-fluence (2.8 mJ/cm$^2$) 1.6-eV case (Fig. 3c) shows a clear signature of IE. In good agreement with Ref. [44], we find from an exponential fit that the number of electrons increases rapidly and thermalizes in $16 \pm 5$ fs. Here, we simultaneously observe the real-time dynamics of Dirac holes, and we find identical behavior, albeit with a slightly faster thermalization time of $12 \pm 3$ fs.

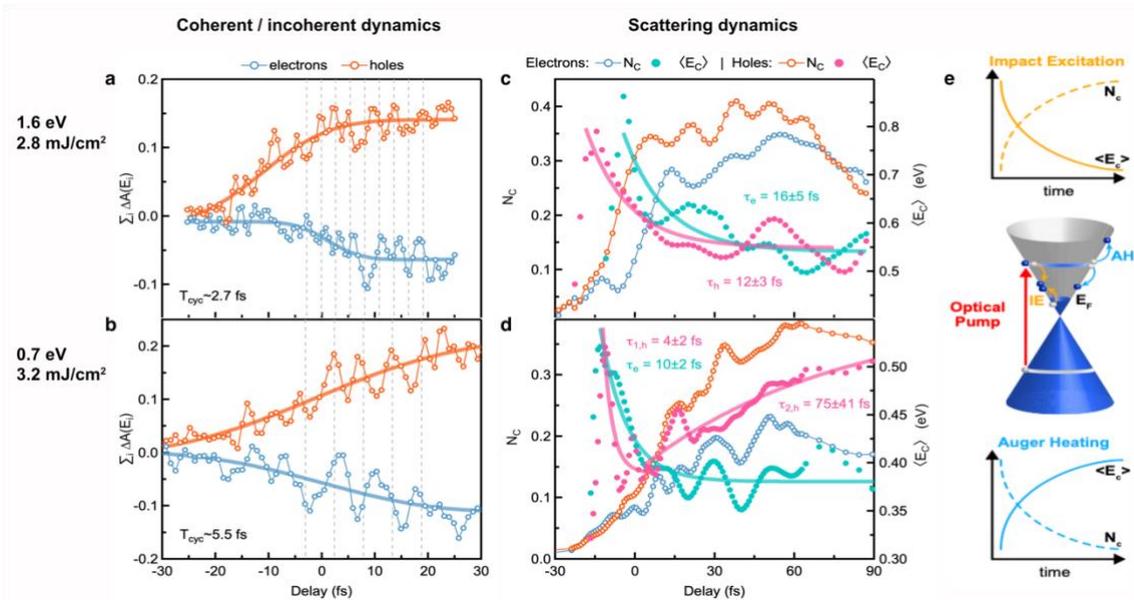

**Fig. 3 Carrier dynamics and identification of scattering processes.** (a) Sum over the differential absorption $\sum_i \Delta A$, which shows the time evolution of electrons (blue) and holes (orange) for a photon energy of 1.6 eV with a fluence of 2.8 mJ/cm$^2$ and delay step size of 0.6 fs. Clearly visible are coherent oscillations at a frequency near the carrier frequency of the light pulse (1.6 eV corresponds to 2.7 fs,

marked as dashed vertical lines) and the buildup of the incoherent background leading to material excitation. (b) shows results for a photon energy of 0.7 eV and pump fluence of 3.2 mJ/cm$^2$. Nicely visible is the buildup of polarization with oscillation frequency of the light carrier wave (0.7 eV corresponds to 5.5 fs, marked as dashed vertical lines) and the material's excitation, albeit at a slower rate. (c,d) Measurements over a larger temporal range showing the evolution of the number of carriers $N_c$ (open symbols) and their average kinetic energy $\langle E_c \rangle$ (closed symbols) for both electrons (blue/petrol) and holes (orange/pink). (e) Illustration of the dominant processes, Auger Heating (AH) and Impact Excitation (IE) on the example of graphene. For both processes, we indicate the expected temporal evolution of $N_c$ and $\langle E_c \rangle$. (c) The scattering mechanism is IE for both carrier types, electrons and holes. (d) Electrons are dominated by IE while the dynamics of holes is governed by IE at the earliest instances of the pump pulse, but then rapidly dominated by AH.

The power of measuring both carrier types in real time becomes obvious when changing to a pump photon energy of 0.7 eV, for similar pump fluence (3.2 mJ/cm$^2$). Figure 3d clearly shows that the measurement of electrons alone would have confirmed IE as a mechanism, while the holes show a more complex behavior. Interestingly, $N_c$ increases comparably for electrons and holes up to 70 fs. However, $\langle E_c \rangle$ behaves differently: Within $10 \pm 2$ fs ($4 \pm 2$ fs for holes), electrons continue to lose energy while holes initially lose energy but then, on average, gain energy. Holes thus exhibit a switchover from IE to AH during the 11-fs-duration of the infrared light field. Considering the lower photon energy of 0.7 eV, which generates electron-hole pairs much closer to the Fermi energy, the previously inferred n-doping of 200 meV leads to an asymmetric phase space for scattering[47]. This asymmetry explains the electron-hole symmetry breaking for scattering in the conduction band (CB) and effectively suppresses hole IE [45,48]. To the best of our knowledge, this is the first direct measurement of the real time dynamics for both carrier types and it shows the importance of such measurement to address problems related to energy storage or light harvesting devices where the recombination of carriers plays a major role in their functionality.

Having elucidated the mechanisms of carrier scattering, we now turn to analyzing the full-time range of our measurement, up to 1 ps, to investigate electron-phonon scattering and the time evolution of phonons together with their dispersion. Thus, we first fitted the incoherent contribution of the signals shown in Fig. 2b to access the coherent dynamics. We employ a three-temperature model rate-equation model (3TM)[49]. This simple model provides a phenomenological thermodynamic description by subsystems of electrons, strongly-coupled optical phonons (SCOPs) and lattice; see the SI for details. The 3TM fit to the data is shown in Fig. 4a and yields that while SCOPs represent only 0.24% of all vibrational modes, they efficiently absorb ~90% of the energy from the electronic sub-system due to the large electron-phonon coupling (EPC) strength of 2.8x10$^{16}$ W/m$^3$K. We note that these values are physically reasonable and in agreement with literature[26].

Next, we investigate the coherent phonon[50,51] signal by analyzing the oscillatory pattern exhibited by the $\sigma^*$ data (Fig. 2b, orange circles and Fig. 4b) with a short-time Fourier transform (STFT) analysis. Such analysis helps visualizing the phonon dispersion landscape (Fig. 4d) and it allows to identify various dominant coherent phonon modes. We note that any STFT naturally requires balancing frequency versus time resolution, we have thus conducted an additional Fourier analysis of the signal over the entire delay range of 1 ps (Fig. 4c); see the SI. Figure 4d shows that already during and shortly after the laser excitation, coherent motion emerges over a broad range of frequencies. To identify these frequencies, we performed large-scale two-temperature-model molecular dynamics (TTM-MD) simulations using an electronic-temperature dependent interatomic potential for graphite (analog to Ref. [52,53]) and the *ab-initio* determined Eliashberg-function and *k*-dependent coupling strength; details are given in the SI. Results from the simulation are shown together with experimental data. Figure 4a shows the time evolution of the temperatures of electrons, Fig. 4b for SCOPs (the latter calculated directly without approximation from the kinetic energy of the corresponding modes) and the lattice. We find that the match with the experiment is remarkable considering that no fitting parameter was used. Figure 4e shows the calculated phonon dispersion for a range of relevant electron temperatures. We determine the two highest phonon frequencies from a multipeak fit to the data and show them in Fig. 4c in blue and red. A comparison with the phonon dispersion (Fig. 4e) identifies them as the Raman-active $\bar{\Gamma} - E_{2g}$ and the non-Raman-active $\bar{K} - A_1'$ SCOPs at 46.4 $\pm$ 2.7 THz and 42.7 $\pm$ 1.1 THz, respectively. We also note the unexpected observation of very high-frequency coherent lattice oscillations in a forbidden range of the phonon spectrum up to 90 THz (see Fig. 4c, top left). These frequencies are commensurate with twice the frequency of the $\bar{K} - A_1'$ phonon, and, to the best of our knowledge, have not been observed in graphite previously. The capacity to simulateneously measure both, the Raman-active $\bar{\Gamma} - E_{2g}$ and the non-Raman-active $\bar{K} - A_1'$ SCOP allows to reveal their temporal dynamics; see the red and blue lineouts across the STFT. Unexpectedly, we find that both SCOPs become dominant already after 20 fs and they reach their maxima at 65.5 $\pm$ 3.0 fs and 69.6 $\pm$ 3.0 fs, respectively. The early onset of coherent oscillations of the $A_1'$ mode is startling, considering that the only coherent phonon that can be excited according to present understanding is the Raman active $E_{2g}$ mode. Equally surprising, the TMM-MD model reveals that the $A_1'$ SCOP provides the dominant pathway (~90%) for deexcitation of the electronic subsystem.

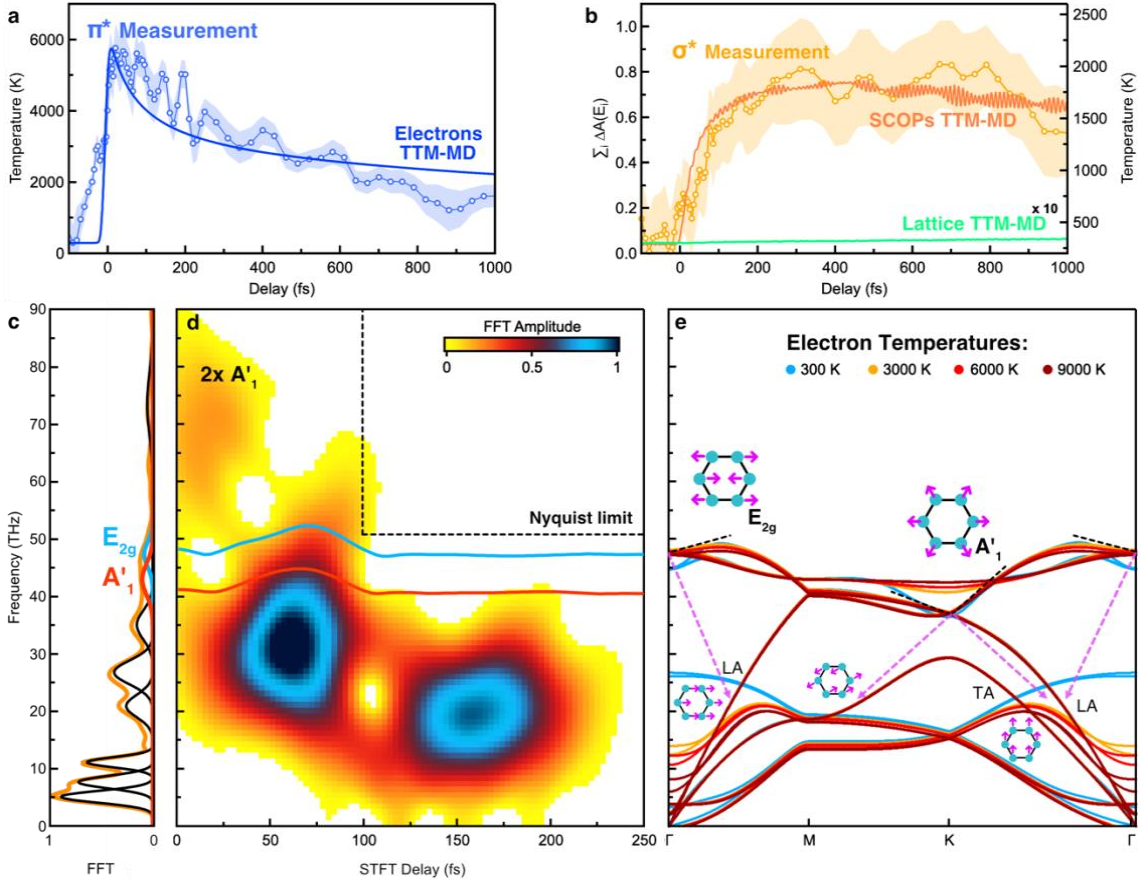

**Fig. 4 Phonon dynamics retrieved from attosecond-XANES.** (a) and (b) show results from the TTM-MD simulation (solid lines) together with experimental data from Fig. 2. The evolution of the electronic temperature is shown in (a) while (b) shows the SCOPs and lattice subsystems. (c) shows results from a Fourier analysis of the experimental $\sigma^*$ data (shown in (b)). A multipeak fit reveals the predominant modes; positions of the $A_1'$ and $E_{2g}$ are shown in red and blues, respectively. (d) Short-Time Fourier Transform (STFT) of the $\sigma^*$ data (b) over a time range of 250 fs, visualizing coherent phonon contributions. A comparison with the calculated phonon dispersion (e), for different electronic temperatures (see SI), allows to identify the phonon

modes, their main branches (dashed lines), and their dispersion (dashed arrows). Striking are the high frequency oscillations at twice the phonon frequency of $A_1'$, the simultaneous and fast rise of the coherent Raman active $E_{2g}$ mode, concomitantly with the non-Raman-active $A_1'$, mode, and the rapid decay into coherent low-frequency phonons and their temporal dynamics.

To elucidate the surprising early contribution from the (non-Raman-active) $A_1'$ mode, we calculated the equilibrium and the laser-excited potential energy surfaces along the $A_1'$ mode (see SI), and we found no displacement of the potential minimum. Thus, without the possibility for an immediate displacive excitation of the fully symmetric mode, we conclude that the observed coherent lattice motion must originate from the very strong electron-SCOPs coupling, thus acting almost impulsively. Our model further explains the unexpected observation of the high frequency oscillations up to 90 THz at twice the frequency of the $A_1'$ mode. We find that the DOS is modulated due to the absolute value of the displacement but not its direction, thus resulting in twice the phonon oscillation frequency in the DOS. The simulations also explain the surprising finding that the oscillations of the $A_1'$ mode are stronger than those of the $E_{2g}$ phonon, and that it reaches its maximum earlier: The form of the Eliashberg function yields much stronger electron-phonon coupling for the $A_1'$ mode than for the $E_{2g}$ mode [53]. Thus, in accord with the experiment, even though the $E_{2g}$ is impulsively excited by the light field via Raman scattering, the main loss channel for electronic de-excitation is via the $A_1'$ mode due to the very strong EPC. This also explains the striking observation that the coherent oscillation amplitudes of both SCOPs reach their maxima ~60 fs after the pump pulse. Further analysis reveals excitation of the $\bar{K} - E'$ mode at 33.7 $\pm$ 0.1 THz, and the observation of lower energetic phonon modes at 26.6 $\pm$ 0.1 THz, 19.8 $\pm$ 0.1 THz and 16.5 $\pm$ 0.1 THz, which peak after ~160 fs. This observation is in agreement with recent results by Stern et al. [54] from ultrafast electron diffraction, and suggestive of parametric difference frequency generation[55,56] from $E_{2g}$ and $A_1'$ phonons into less energetic coherent transverse and longitudinal acoustic phonon modes.

In conclusion, we demonstrate the ability to track the flow of energy upon light absorption between electrons, holes, and phonons simultaneously and in real time. Our method expands on K-shell core-level XANES spectroscopy with attosecond temporal resolution to achieve an unambiguous and simultaneous view on the temporal evolution of the photon-carrier-phonon system. A carrier-envelope-phase (CEP) stable pump pulse enables the attosecond probe to resolve the buildup of electronic coherence (polarization) even during the excitation light field's oscillation at PHz oscillation frequency. The broad bandwidth of the attosecond SXR pulse provides concomitant spectral coverage over a range of several 100's eV. Notably, the measurement's energy resolution is limited only by the x-ray spectrometer and not by choice of the pump-probe measurement's time steps. Applying this methodology to graphite, the ability to simultaneously measure the dynamics of electrons and holes reveals disparate scattering mechanisms for infrared excitation, close to the Fermi energy. We find that ultrafast dephasing of the coherent carrier dynamics is governed by Impact Excitation for electrons, while holes exhibit a switchover from Impact Excitation to Auger heating during the 11-fs-duration of the infrared light field. We attribute this switchover to the limited scattering phase space in the n-doped material. Further, the measurement sheds new light on long-standing questions regarding the excitation of SCOPs in graphite. We directly observe the non-Raman active $A_1'$ phonon concomitantly with the Raman-active $E_{2g}$ phonon, and we find that both SCOPs are coherently excited within 20 fs of the pump pulse. The coherent excitation of both SCOPs is non-displacive and is explained by the strong electron-phonon scattering, i.e., via a seemingly incoherent process. We identify the $A_1'$ phonon as the dominating channel for dissipation of electronic coherence. Moreover, unobserved in graphite, we find high-frequency oscillations up to 90 THz, which arise from the modulation of the electronic DOS by the atomic displacements along the $E_{2g}$ and $A_1'$ modes.

These measurements show the utility of our detection methodology even for a seemingly well-studied system like graphite. The method is generally applicable to molecules, liquids and solids and the range of elemental absorption edges that are presently in reach give access to a wide variety of important scientific problems. We thus expect it may prove valuable to address questions such as, for instance, the energy dissipation in light-harvesting, organic electronic and energy storage systems, or to re-examine long-standing questions in non-equilibrium multi-body physics such as phase-transitions and superconductivity. The ongoing development of x-ray light sources, either high harmonic-based or free electron lasers, will bring the methodology to elemental K-shell edges of heavier elements in the hard x-ray spectral range.

**Acknowledgements**

J.B. acknowledges financial support from the European Research Council for ERC Advanced Grant "TRANSFORMER" (788218), ERC Proof of Concept Grant "miniX" (840010), FET-OPEN "PETACom" (829153), FET-OPEN "OPTOlogic" (899794), Laserlab-Europe (871124), Marie Sklodowska-Curie ITN "smart-X" (860553), MCIN for PID2020-112664GB-I00 (AttoQM); AGAUR for 2017 SGR 1639, MINECO for "Severo Ochoa" (SEV- 2015-0522), Fundació Cellex Barcelona, the CERCA Programme / Generalitat de Catalunya, and the Alexander von Humboldt Foundation for the Friedrich Wilhelm Bessel Prize. M.R. acknowledges Marie Sklodowska-Curie grant agreement 754510 (PROBIST). S. Sharma thanks DFG for funding through TRR227 (project A04). M.E.G. acknowledges support from DFG through grant GA465/15-2. The large-scale MD simulations were performed on the Lichtenberg High Performance Computer (HHLR) at the TU Darmstadt. We thank J. Menino for his technical support.